\documentclass[%
 reprint,
 amsmath,amssymb,
 aps,
prl
]{revtex4-1}

\usepackage{array,url,kantlipsum}
\usepackage{soul}
\usepackage{color}
\usepackage{graphicx}
\usepackage{dcolumn}
\usepackage{bm}

\begin{document}

\preprint{APS/123-QED}

\title{Frictional state evolution laws and \\ the non-linear nucleation of dynamic shear rupture}

\author{Robert C. Viesca}

\affiliation{Department of Civil and Environmental Engineering, Tufts University, Medford, MA 02155 USA}
\date{\today}

\begin{abstract}

We assess if a characteristic length for a non-linear interfacial slip instability follows from theoretical descriptions of sliding friction. We examine friction laws and their coupling with the elasticity of bodies in contact and show that such a length does not always exist. We consider a range of descriptions for frictional strength and show that the area needed to support a slip instability is negligibly small for laws that are more faithful to experimental data. This questions whether a minimum earthquake size exists and shows that the nucleation phase of dynamic rupture contains discriminatory information on the nature of frictional strength evolution. 

\end{abstract}

\maketitle
\vspace*{-1.25 cm}
\section{1. Introduction}

Frictional interfaces of sufficiently compliant bodies allow for one part of the interface to slip while another remains stuck. When interfacial strength weakens with slip, the interplay between friction and elasticity can give rise to quasi-static instabilities, in which slip locally accelerates. However, is there a critical size of an interface, below which a slip instability cannot emerge? Such a question arises when considering whether earthquakes, generated by the frictional rupture of a geological fault, have a minimum size. Observationally in the field, earthquake magnitudes show an ever-decreasing trend as data quality and detection methods are improved \cite{RYDELEK:1989vd,Wiemer:2000vd,Boettcher:2009jm,McLaskey:2014ku,Ross:2019ih}. In contrast, laboratory friction experiments frequently show that early stages of sample-scale seismic rupture occur within a region of finite size \cite{Rubinstein:2004eka,Nielsen:2011hh,Latour:2013bo,Svetlizky:2016gs,Ke:2018ky,McLaskey:2019dp}.

Previous theoretical analyses of nucleation indicated that the process occurs over a finite length. Nucleation length scales have been derived by a linear stability analysis of elasto-frictional coupling \cite{Rice:1983tw, Rice:2001vq,Aldam:2017bia}, analysis of instability progression in a non-linear regime \cite{Dascalu:2000jp,Uenishi:2003gy,Rubin:2005gm,Ampuero:2008bx,Viesca:2016bga,Viesca:2016kua}, or by examining the stability of traveling wave solutions for interfacial rupture \cite{Brener:2018ip}. These analyses rest upon a range of constitutive descriptions for sliding friction that remain phenomenologically posited on laboratory friction experiments. These experiments probe frictional strength by considering its response to a variable history of sliding, with constitutive relations expressed as a dependence on state variables \cite{Dieterich:1978ub,Dieterich:1979ek,Ruina:1983th,Heslot:1994uc}. Examinations of existing friction laws against experimental data has indicated that one description is best able to reproduce observations and is one in which slip is required for frictional strength to evolve \cite{Bhattacharya:2015jw,Bhattacharya:2017fy}. However, this so-called slip law is the one for which theoretical understanding of a nucleation process is the most incomplete.

Here we derive an analytical understanding for nucleation spanning a range of frictional laws and show that length scales associated with nucleation are vanishingly small for the most realistic strength descriptions, implying that there is no theoretical lower limit to the sizes of a slip instability and the concomitant earthquake. To do so, we make use of  a parameterization of frictional strength in which a single parameter allows to span a range of friction laws and the slip law is retrieved for small values of the parameter. This permits a theoretical analysis of the realistic slip law, which has heretofore appeared intractable. We find that the characteristic nucleation length becomes infinitesimally small for slip-law frictional evolution. Furthermore, we find that the nucleation process may itself be unstable and explain the conditions for the emergence of a propagating instability.

An elementary expression for a slip rate and state dependence of the sliding friction coefficient $f$ follows \cite{Ruina:1983th}
\begin{equation}
f(V,\theta) = f_o +a \ln \frac{V}{V_o} + b \ln \frac{\theta}{\theta_o}
\label{eq:f}
\end{equation}
in which $V$ and $\theta$ are the instantaneous sliding rate and a state variable at a point on the interface, and $f_o$, $V_o$, $\theta_o$ are reference values of the friction coefficient, slip rate, and state, respectively. The explicit dependence on sliding rate in (\ref{eq:f}), initially motivated by experimental observation \cite{Dieterich:1978ub,Dieterich:1979ek,Ruina:1983th}, follows from the presumption that creep of asperity contacts under shear is an Arrhenius-activated process \cite{Baumberger:1999wt, Rice:2001vq, Nakatani:2001ut}. The last term captures the dependence of the friction coefficient on the history of sliding. The specific form of the final term---here logarithmic, following convention---is arbitrary.  The importance of the state variable is wholly contained in the equation for its evolution. Evolution laws were proposed along with (\ref{eq:f}) on the basis of the relative weightings of the slip-rate history and experimental data \cite{Ruina:1983th}. Two of these evolution laws, commonly referred to as the aging and slip laws, respectively $\partial \theta/\partial t =1-V\theta/D_c$ and $\partial \theta/\partial t=-V\theta/D_c\ln (V\theta/D_c)$, respect salient features of laboratory friction experiments, in which histories of sliding rate are imposed and the evolving frictional resistance measured. The length $D_c$ is a characteristic slip distance over which state evolution occurs. This description of friction has the property that the rate-strengthening owed to the slip-rate dependence in (1) can be overcome by weakening provided by state evolution, provided the coefficients modulating their relative importance in (\ref{eq:f})---$a$ and $b$---are such that $a<b$.

A linear stability analysis of uniform sliding along an interface of two elastic continuum, in which an infinitesimal perturbation of sliding rate of fixed wavelength is imposed onto an interface uniformly sliding at steady-state, revealed a critical wavelength above which perturbations grow  \cite{Rice:1983tw, Rice:2001vq,Aldam:2017bia}. The critical wavelength is independent of the choice of aging or slip state evolution law, owed to their common linearization about steady state. The critical wavelength's existence has been taken to suggest that a minimum interfacial area is required to initiate an instability of the sliding rate. Further analysis and numerical solutions of interfacial slip with the aging-law (2) has indicated that a characteristic nucleation length may persist into the non-linear regime \cite{Rubin:2005gm,Viesca:2016bga,Viesca:2016kua}. Specifically, under the aging law there exist solutions for slip rate that diverge quasi-statically at a finite time $t_o$ as
\begin{equation}
V(x,t)=\frac{D_c}{t_o-t}\mathcal{W}(x)
\label{eq:div}
\end{equation}
where a localized distribution of slip rate $\mathcal{W}$ has compact support on $|x|<L$ and its spatial distribution depends uniquely on the ratio $a/b$. Diverging slip rate is eventually limited by inertia, which transitions localized acceleration to an outwardly propagating dynamic rupture. $L$ can be considered a nucleation length as it is the characteristic lengthscale of the slip rate distribution just preceding the onset of inertial effects.

However, the relevance of a finite nucleation length based on an aging-law description has been called into question given that the slip law explains a wider range of experimental observations. Measurements of the response of frictional resistance to step changes in sliding rates show a systematic symmetry to step increases or decreases for changes up to three orders of magnitude \cite{Dieterich:1981ue,Tullis:1986ui,Bhattacharya:2015jw}. Model fits using several state evolution laws show that the aging law cannot fit both increases and decreases in sliding rate whereas the slip law provides a robust fit \cite{Bhattacharya:2015jw}. Other probes of frictional strength evolution, including strength response to abrupt halts of sliding, followed by resliding (so-called slide-hold-slide tests), likewise show relatively favorable explanation by the slip law \cite{Bhattacharya:2017fy}.

There are several theoretical indications that a consequence of slip-law-like behavior is that dynamic-rupture-nucleating instabilities may develop over negligible distances along an interface. Non-linear stability analysis of a single-degree-of-freedom spring-loaded sliding block model shows, for the slip law, that a slip-rate instability can occur with any spring stiffness, provided perturbations are sufficiently large \cite{Gu:1984ua}. This result can be extended to an interface of continuous bodies, in which stiffness of the spring-block system can be inversely related to the spatial wavelength of a continuous monochromatic perturbation, implying that arbitrarily short wavelength perturbations may be unstable. This contradicts the linear stability analysis, which suggested short-wavelength perturbations are stable, and lies in contrast to a similar non-linear analysis with the aging law that shows unconditional stability below a critical wavelength \cite{Ranjith:1999wv}. Furthermore, numerical solutions of sliding interfaces of elastic continua with the slip law show the acceleration of a fault towards instability occuring over ever-diminishing length scales; these same solutions also exhibit accelerating slip propagating as a slip pulse, something not observed under the aging law \cite{Ampuero:2008bx,Rubin:2009dk}. 

\begin{figure}
\includegraphics{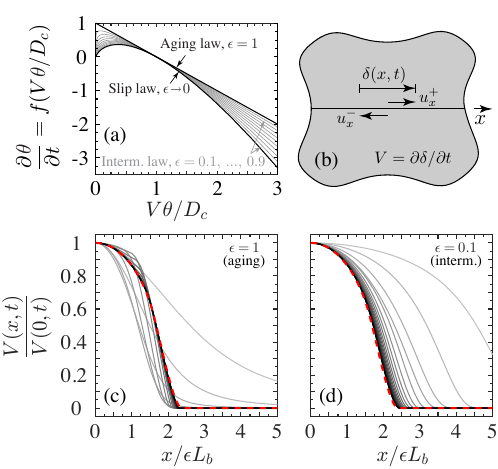}
\caption{\label{figdiv}(a) Illustration of one-parameter family of state evolution laws for sliding friction, (\ref{eq:interm}). (b) Definition of slip $\delta$ and slip rate $V$ for an example of in-plane distribution of relative interfacial displacement. (c,d) Results from numerical solutions for slip-rate evolution for rupture of a rate-weakening interface ($a<b$) at late stages of quasi-static instability. The spatial distribution of a diverging slip rate is shown at instants in time, with time progress corresponding to darkening greyscale. Slip rate is scaled by its instantaneous value at $x=0$, which is diverging and about which the distribution is symmetric. Distance $x$ is scaled by $\epsilon L_b$ where $L_b$ is an elasto-frictional length scale (Appendix A.3). Solutions are found for different values of parameter $\epsilon$ and the same value of parameter $a/b=0.6$. The slip rate in (d) diverges over distances ten-times smaller than that in (c). The solutions approach the same distribution (red-dashed) as the instability progresses.}
\end{figure}

\section{2. Instability analysis using an intermediate state evolution law}

We analyze the spatiotemporal development of accelerating slip into the non-linear regime on an interface whose frictional strength evolution follows the slip law. To do so, we find a convenient route via an intermediate state evolution law \cite{intermref}
\begin{equation}
\frac{\partial\theta}{\partial t}= \frac{1}{\epsilon}\left[\left(V\theta/D_c\right)^{-\epsilon}-1\right]V\theta/D_c
\label{eq:interm}
\end{equation}
from which we retrieve the aging law for $\epsilon=1$ and the slip law in the limit $\epsilon\rightarrow 0$. 

We find that solutions for diverging slip rate in the form (\ref{eq:div}) exist for any value of $\epsilon$ in this range, provided $\mathcal{W}(x)$ in (\ref{eq:div}) is replaced with $\epsilon \mathcal{W}(x/\epsilon)$ (Appendix A): i.e., distances and the nucleation length $L$ are scaled down by a factor $\epsilon$. In Figure \ref{figdiv}, we demonstrate this result by showing the slip rate evolution along an interface at snapshots in time as an instability develops, under different values of the parameter $\epsilon$. The numerical solutions model the accelerating in-plane (or anti-plane) rupture of an interface of two linearly elastic half-spaces (Appendix C.1). As time progresses, the scaled slip rate asymptotically approaches the same distribution, $\mathcal{W}(x/\epsilon)$. The vanishing nucleation length in the slip-law limit of $\epsilon\rightarrow 0$ implies that pointwise divergence of slip rate may be possible. The result is surprising for a system whose linear stability analysis indicated elastic interactions would tend to delocalize such pointwise divergence. 

In addition to confirming our expectations using numerical solutions, we independently assess the stability of the diverging slip-rate solutions (\ref{eq:div}) for any value of the parameter $\epsilon$. Specifically, we perform a linear stability analysis of the non-linear self-similar solutions (\ref{eq:div}) \cite{Barenblatt:1996ea,Eggers:2015ux}. To do so, we define an alternate pair of variables to slip rate $V$ and state $\theta$, which we denote as $W$ and $\Phi$. The first is defined implicitly by $V(x,t)=\epsilon W[x,t(s)] D_c/(t_o-t)$ and corresponds to a possible variation with time of the spatial distribution $\mathcal{W}$. The change of independent variable from $t$ to $s$ follows $s=-\log(t_o-t)$, which approaches infinity as $t\rightarrow t_o$. The second alternate variable is defined $\Phi(x,s)=1-(V[x,t(s)]\theta[x,t(s)]/D_c)^{-\epsilon}$ and is chosen for convenience. The stability analysis determines whether we should expect the distribution $W(x,s)\rightarrow\mathcal{W}(x/\epsilon)$ as $s\rightarrow \infty$, and for the distribution $\Phi$ to approach an analogous distribution $\mathcal{P}(x/\epsilon)$. 

We linearize the evolution equations for small perturbations of $W$ and $\Phi$ about $\mathcal{W}$ and $\mathcal{P}$. That is we look for the behavior of perturbations of the form $W(x,s)=\mathcal{W}(x)+w(x,s)$ and $\Phi(x,s)=\mathcal{P}(x)+\phi(x,s)$. The linearized evolution equations for $w$ and $\phi$ comprise an autonomous system of the form (Appendix B.1)
\begin{align}
\frac{\partial w}{\partial s}&=g(w,\phi)\notag\\[2 pt]
\frac{\partial \phi}{\partial s}&=\epsilon h(w,\phi) \label{eq:sys}
\end{align}
which, in the limit of small $\epsilon$, represents a slow-fast dynamical system. To determine the stability of the  solutions (\ref{eq:div}), we look for solutions to this system of the form $w(x,s)=\omega(x)\exp(\lambda s)$ and $\phi(x,s)=\varphi(x)\exp(\lambda s)$. Given the linear dependence of $g$ and $h$ on their arguments, the problem reduces to a linear eigenvalue problem with eigenvalues $\lambda$ and eigenmodes $\omega$ and $\phi$. 

\begin{figure}
\includegraphics{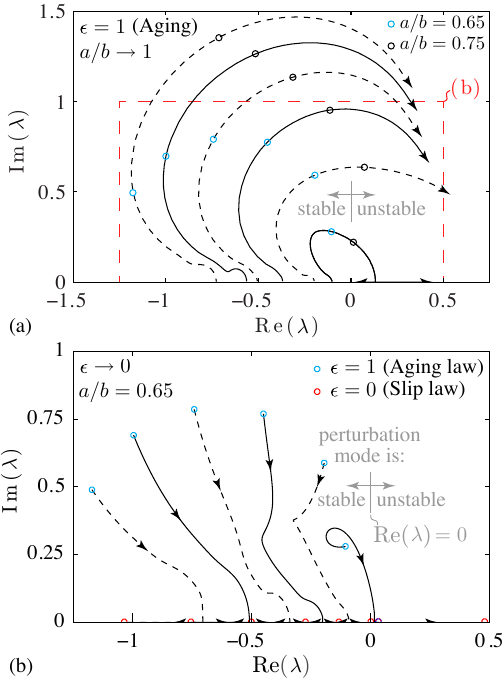}
\caption{\label{figeig}Results from linear stability analysis of diverging slip-rate solutions (\ref{eq:div}). Illustration of eigenvalue trajectories in complex plane (black curves) as (a) $\epsilon$ is fixed and $a/b$ is varied, and (b) $a/b$ is fixed and $\epsilon$ is varied. (a) At fixed $\epsilon$, loss of stability occurs via Hopf bifurcations $a/b$ is increased \cite{Viesca:2016bga}. Open circles are eigenvalue positions for the six least stable modes at fixed values of $a/b$. (b) At fixed $a/b$, a stable blow-up solution (\ref{eq:div}) becomes unstable and the eigenvalues tend towards a purely real set (red circles) as the state evolution law transitions from aging to slip ($\epsilon\rightarrow0$). }
\end{figure}

We assess the stability of the finite-time divergence of interfacial slip rate (\ref{eq:div}) for any value of the two dimensionless parameters, $\epsilon$ and $0<a/b<1$. The first parameter indicates the manner of state evolution and the second parameter indicates the degree of steady-state rate-weakening. Solutions are found to be unstable by solving the aforementioned eigenvalue problem and determining the parameter range for which $\text{Re}(\lambda)>0$. Prior work established that, for the aging law ($\epsilon=1$), the solutions (\ref{eq:div}) lose stability by a cascade of Hopf bifurcations as the parameter $a/b$ increases \cite{Viesca:2016bga}. The Hopf bifurcations' trajectories as pairs of eigenvalues crossing the imaginary axis is shown in Figure \ref{figeig}a, in which the trajectories are symmetric about $\text{Im}(\lambda)=0$.

\begin{figure}
\includegraphics{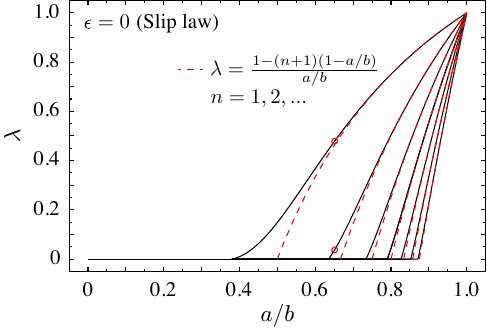}
\caption{\label{figreeig} Numerical solutions (black) for eigenvalues $\lambda\geq0$. Pointwise divergence is a stable, attractive solution for ${a/b<0.3781...}$ Above this critical value, unstable modes appear sequentially. Eigenvalues for the first seven unstable modes are shown, with asymptotic behavior as $a/b\rightarrow 1$ (red-dashed). Two positive eigenvalues at $a/b=0.65$ (red circles) shown for comparison with Figure \ref{figeig}b.}
\end{figure}

To determine the stability of slip-law blow-up solutions in the form (2), we now fix the value of $a/b$ and vary $\epsilon$ from its aging-law to slip-law end-member values (i.e., decreasing $\epsilon$ from 1 to 0). We begin with a value of $a/b=0.65$, for which solutions (2) are stable for the aging law. We find that the complex eigenvalues converge to to a purely real set (red circles) as $\epsilon\rightarrow 0$ (Figure \ref{figeig}b). We make use of the slow-fast structure of the system (\ref{eq:sys}) to find the real eigenvalues in this limit (Appendix B.2). In Figure \ref{figreeig} we show the emergence of unstable modes for the slip law ($\epsilon=0$) when the rate-weakening parameter $a/b$ is increased. 

\section{3. Emergence of instability propagation}
How is the stability or instability of solutions (\ref{eq:div}) exhibited? For the aging law, unstable modes have complex eigenvalues (Figure \ref{figeig}a). Examining the evolution of slip rate with $\epsilon=1$ (aging law), an instability occurs as a stable attraction to slip accelerating with the compact, locally peaked manner of (\ref{eq:div}), for $a/b$ below a critical value. Increasing $a/b$, unstable modes emerge and lead to limit-cycle, apparent quasi-periodic, or chaotic oscillations about the solution (\ref{eq:div}) \cite{Viesca:2016bga,Viesca:2016kua}. For the slip law, instability occurs as an accelerating distribution of slip that continuously contracts towards a point when $a/b<0.3781...$, for in-plane or anti-plane rupture of two half-spaces.  Unlike the aging law, this pointwise divergence loses stability with a strictly real set of eigenvalues for $a/b>0.3781...$ . This leads to an absence of limit-cycle or otherwise aperiodic oscillations of accelerating slip about a single point. This also explains prior numerical solutions under the slip law, which show slip acceleration occuring as a unilateral slip pulse \cite{Ampuero:2008bx,Rubin:2009dk}: a solution exists that permits blow-up at a point, but any point in space is equally unattractive. Prior work sought to explain the slip-pulse emergence by excluding the possibility of localized acceleration in a crack-like manner, using heuristic arguments regarding the scaling of apparent fracture energy for the slip law in response to rapid jumps in slip velocity \cite{Ampuero:2008bx}.

\begin{figure}
\includegraphics{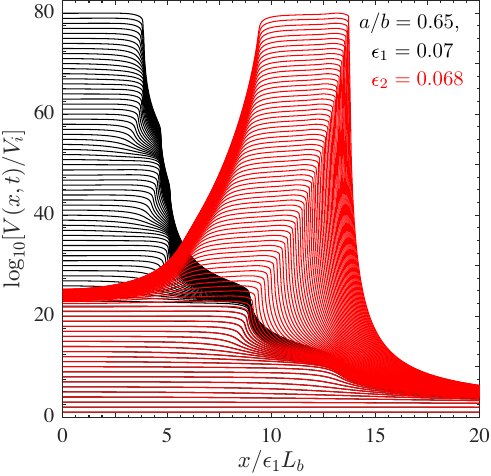}
\caption{\label{fignum} Snapshots of the distribution of slip velocity $V$ over positions along the interface $x$ at instants in time for a fixed value $a/b=0.65$ and two values of $\epsilon$ near a stability transition point. Time snapshots are at equal intervals of the peak slip rate. As the slip law is approached with decreasing $\epsilon$, an abrupt transition from localized to migrating slip occurs for such large values of $a/b$.}
\end{figure}

In Figure \ref{fignum}, we highlight the transition from localization to migration of accelerating slip as a slip-law state evolution is approached.  We follow the instability progression for a fixed value of $a/b$ and two values of $\epsilon$ close to the stability transition point in Figure \ref{figeig}b, in which an eigenvalue trajectory for a single mode crosses the imaginary axis. For $\epsilon=0.07$, localized acceleration is attractive and occurs over a finite, but small nucleation length, $L\approx4\epsilon L_b$. As $\epsilon$ is decreased slightly to 0.068, the localized acceleration loses stability and transitions to a migrating pulse of accelerating slip. For $a/b<0.3781...$, the localized acceleration would remain distributed about the point $x=0$, but would occur over vanishingly small distances along the interface as $\epsilon\rightarrow 0$. 

Inertia ultimately limits unstable quasi-static acceleration. Prior numerical solutions incorporating first-order inertial effects, in the form of the radiation damping, show the principal features of the non-linear solutions (\ref{eq:div}) and their stability in the moments leading up to dynamic rupture initiation, for both aging and slip laws \cite{Rubin:2005gm,Rubin:2009dk}.

\section{4. Conclusion}
To summarize, we examined the local, quasi-static acceleration of slip on an interface, which initiates the transition from a stuck to sliding interface by way of the nucleation of a dynamic rupture. We provided a theoretical analysis for the non-linear instability development for a frictional strength obeying a slip-rate and state-variable dependence, over a range of previously proposed state evolution laws. We found non-linear solutions for slip-rate divergence over this range and that the associated nucleation length scale vanishes as the slip law of state evolution is approached. This implies that dynamic ruptures may initiate over distances much smaller than previously suggested by linear stability analysis of interfacial slip or by non-linear behavior of frictional laws that are less capable of reproducing experimental data. 

What are the conditions for the initiation of an instability over such small regions? Spring-block model results show a finite perturbation from steady state is required for an instability to emerge in system whose spatial length is below a critical wavelength \cite{Gu:1984ua}. In continuum systems, a finite perturbation can initiate a slip instability over distances smaller than would be expected from linear stability analysis (Appendix C.2). Finite perturbations can follow from a combination of the heterogeneous distribution of slip due to a history of events combined with an external forcing. The contrast of locked and sliding regions along an interface have been observed to lead to reduced dynamic-rupture nucleation lengths in experiments on spatially extended frictional interfaces, when creeping fronts converge on a stuck asperity, leading to sudden, localized increases in loading rate \cite{McLaskey:2019dp}.

\begin{acknowledgments}
This work was supported by the National Science Foundation (grants EAR-1344993 and EAR-1834696).
\end{acknowledgments}

\bibliography{bibliography}

\begin{thebibliography}{36}
\bibitem [{int()}]{intermref}%
  \BibitemOpen
\bibfield  {journal} {  }\href@noop {} {\bibinfo  {journal} {Derived
  independently for the purposes used here, this intermediate law was also
  found to be originally presented in [23]}\ }\BibitemShut {NoStop}%
\end{thebibliography}

\pagebreak


\setcounter{figure}{0}
\renewcommand{\thefigure}{C.\arabic{figure}}

\section{Appendix A}
\setcounter{equation}{0}
\renewcommand{\theequation}{A.\arabic{equation}}

\subsection{A.1 Evolution equations}
We derive the evolution equations for the slip rate $V$ and an alternative state variable $\Phi$. The evolution equations come from an empirical evolution equation for a state variable $\theta$ and the requirement that the frictional strength $\tau_s$ equals the shear stress $\tau$ on a sliding interface. This latter equation we may write as 
\begin{equation}
\sigma f=\tau_b+\mathcal{L}(\delta)
\label{eq:start}
\end{equation}
where $\sigma$ is the interface-normal component of stress, $\tau_b$ is the shear stress resolved on the interface without any slip and $\mathcal{L}(\delta)$ is the change in interfacial shear stress due to the elastic response of bodies having a distribution of slip $\delta$ along their interface. For the in-plane or anti-plane sliding of a thin layer on a substrate, or such sliding contact of two half-spaces, the operator $\mathcal{L}$ takes the form, respectively,
\begin{equation}
\mathcal{L}(\delta)=\bar E h \partial \delta^2/\partial x^2, \quad \mathcal{L}(\delta)=\bar \mu \mathcal{H}(\partial \delta/\partial s)
\label{eq:L}
\end{equation}
where $\mathcal{H}$ is the Hilbert transform, $\bar \mu$ and $\bar E$ are sliding-mode-dependent moduli, and $h$ is the layer thickness. For anti-plane and in-plane sliding, respectively, $\bar \mu$ equals $\mu$ and $\mu/(1-\nu)$ and $\bar E$ equals $\mu$ and $2\mu/(1-\nu)$.

For a bounded, negligible rate of external loading, the time derivative of (\ref{eq:start}) is
\begin{equation}
\sigma\left(\frac{a}{V}\frac{\partial V}{\partial t}+\frac{b}{\theta}\frac{\partial\theta}{\partial t}\right)=\mathcal{L}(V)
\end{equation}
Noting that the intermediate state evolution evolution law (3) may be written in terms of $V$ and $\Phi$ as
\begin{equation}
\frac{1}{\theta}\frac{\partial \theta}{\partial t}=-\frac{V\Phi}{D_c\epsilon}
\label{eq:evolTh}
\end{equation}
the equation for the evolution of slip rate is
\begin{equation}
\frac{\partial V}{\partial t}=\frac{b}{a}\left(\frac{V\mathcal{L}(V)}{\sigma b}+\frac{V^2\Phi}{\epsilon D_c}\right)
\label{eq:evolV}
\end{equation}
We may derive the evolution equation for $\Phi$ by taking the time derivative of its definition (3)
\begin{equation}
\frac{\partial \Phi}{\partial t}=\epsilon (1-\Phi)\left(\frac{1}{V}\frac{\partial V}{\partial t}+\frac{1}{\theta}\frac{\partial\theta}{\partial t}\right)
\end{equation}
and after substituting (\ref{eq:evolTh}) and (\ref{eq:evolV}), we find
\begin{equation}
\frac{\partial \Phi}{\partial t}=\epsilon \frac{b}{a}(1-\Phi)\left(\frac{\mathcal{L}(V)}{\sigma b}+\left(1-\frac{a}{b}\right)\frac{V\Phi}{\epsilon D_c}\right)
\label{eq:evolPhi}
\end{equation}

\subsection{A.2 Change of variables}
Following a scaling analysis, we find that similarity solutions may exist and that to search for these solutions, an appropriate change of dependent and independent variables is \cite{Barenblatt:1996ea, Eggers:2015ux}
\begin{equation}
V(x,t)=\epsilon D_c \frac{W[x,s(t)]}{t_o-t},\quad \frac{ds}{dt}=\frac{1}{t_o-t}
\end{equation}
where we also perform the change of dependent variable $\Phi(x,t)\Rightarrow\Phi(x,s)$. With the anticipated finite-time divergence of slip rate $V$, we are now no longer explicitly interested in the evolution of slip rate $V$ and the original state variable $\theta$ themselves, but rather the quantities $W$ and $\Phi$. Starting from (\ref{eq:evolPhi}), we may write the evolution of $\Phi$ with respect to the new dependent variable $s$ as
\begin{equation}
\frac{\partial \Phi}{\partial s}=\epsilon\frac{b}{a}(1-\Phi)\left(\frac{\epsilon D_c\mathcal{L}(W)}{\sigma b}+\left(1-\frac{a}{b}\right)W \Phi\right)
\label{eq:evolPhis}
\end{equation}
Similarly, beginning with (\ref{eq:evolV}), the evolution equation for $W$ is 
\begin{equation}
W+\frac{\partial W}{\partial s}=\frac{b}{a}\left(\frac{\epsilon D_c\mathcal{L}(W)}{\sigma b}W+W^2 \Phi\right)
\label{eq:evolW}
\end{equation}

\subsection{A.3 Elastofrictional lengthscales}
An  elasto-frictional lengthscale emerges in this problem. The lengthscale for the sliding of a thin layer, or from the whole bodies in contact are, respectively, \cite{Rubin:2005gm,Viesca:2016bga}
\begin{equation}
L_{bh}=\sqrt{\bar E h D_c/(\sigma b)}, \quad L_b=\bar \mu D_c/(\sigma b) 
\end{equation}
That such lengthscales exist is evident from the products involving the elastic operator $\mathcal{L}$ in (\ref{eq:evolPhis}) and (\ref{eq:evolW}). We may simplify these products in the following manner
\begin{equation}
\frac{\epsilon D_c}{\sigma b}\mathcal{L}(W)=\hat{\mathcal{L}}(W)
\end{equation}
where in doing so, the operator $\hat{\mathcal{L}}$ is understood to have had distances scaled by a factor $\epsilon L_b$ or by $\sqrt{\epsilon}L_{bh}$. This is apparent, for example, by considering the thin-layer problem, for which
\begin{equation}
 \frac{\epsilon D_c}{\sigma b}\bar E h\frac{\partial ^2 W}{\partial x^2} = \frac{\partial ^2 W}{\partial (x/\sqrt{\epsilon} L_{bh})^2}
\end{equation}

\

\subsection{A.4 Non-linear similarity solutions for diverging slip rate}
We now look for similarity solutions for which $W(x,s)=\mathcal{W}(x)$ and $\Phi(x,s)=\mathcal{P}(x)$.
Substituting this into (\ref{eq:evolPhis}) and (\ref{eq:evolW}) and combining the results, leads to the equations governing $\mathcal{W}$ and $\mathcal{P}$
\begin{align}
\frac{a}{b}&=\mathcal{W}\mathcal{P}+\tilde{\mathcal{L}}(W)\\
0&=(1-\mathcal{P})(1-\mathcal{W}\mathcal{P})
\end{align}
which are exactly the same equations governing the similarity solutions for the aging law ($\epsilon =1$), except here distances are scaled by an $\epsilon$-dependent elasto-frictional lengthscale. The solutions for the functions $\mathcal{W}(x)$ and $\mathcal{P}(x)$ have a single parameter, $a/b$, and have been found numerically or in closed form \cite{Viesca:2016bga,Viesca:2016kua}.

\section{Appendix B}
\subsection{B.1 Linear stability analysis of similarity solutions}
We assess the stability of these similarity solutions by examining the behavior of small perturbations. If the solutions are stable, then slip rate will be attracted to diverging in the manner of the similarity solutions. We denote the perturbations $w$ and $\phi$
\begin{align}
W(x,s)=&\mathcal{W}(x)+w(x,s)\notag\\ \Phi(x,s)=&\mathcal{P}(x)+\phi(x,s)
\end{align}
Substituting the above into (\ref{eq:evolPhis}) and (\ref{eq:evolW}), we find that, to linear order, the evolution equations for these perturbations are
\begin{equation}
\frac{\partial w}{\partial s}=\frac{b}{a}\left(\mathcal{W}\tilde{\mathcal{L}}(w)+\mathcal{W}\mathcal{P}w\right) +\frac{b}{a}\mathcal{W}^2\phi
\label{eq:evolwpert}
\end{equation}
and
\begin{align}
\frac{\partial \phi}{\partial s}=&\epsilon \frac{b}{a}(1-\mathcal{P})\left(\tilde{\mathcal{L}}(w)+\left(1-\frac{a}{b}\right)(\mathcal{W}\phi+\mathcal{P}w)\right)\notag\\&-\epsilon \phi(1-\mathcal{W}\mathcal{P})
\label{eq:evolphipert}
\end{align}

Looking for solutions in the form $w(x,s)=\omega(x)\exp(\lambda s)$ and $\phi(x,s)=\varphi(x)\exp(\lambda s)$, the problem reduces to one of solving numerically for eigenvalues $\lambda$ corresponding to eigenmodes $\omega$ and $\varphi$, given a choice of parameters $\epsilon$ and $a/b$, the last of which also determines the functions $\mathcal{W}$ and $\mathcal{P}$ (see preceding section). The results of these numerical solutions are presented in Figures 2, 3 as black curves and red circles.

\subsection{B.2 Asymptotic behavior of eigenvalues, eigenmodes}
The slow-fast nature of the pair of evolution equations (\ref{eq:evolwpert})--(\ref{eq:evolphipert}), in which every term in (\ref{eq:evolphipert}) is preceded by $\epsilon$, allows for a simplification of the eigenvalue problem. In the limit $\epsilon\rightarrow 0$, we neglect the evolution of the perturbation $\phi$ and (\ref{eq:evolwpert}) reduces to
\begin{equation}
\frac{\partial w}{\partial s}=\frac{b}{a}\left(\mathcal{W}\tilde{\mathcal{L}}(w)+\mathcal{W}\mathcal{P}w\right)
\label{eq:redeig}
\end{equation}
We may solve this reduced problem in the limit $a/b\rightarrow 1$, for which the product $\mathcal{WP}=1$ on $|x|<L$ \cite{Viesca:2016bga}. Furthermore, for a specific choice of elastic configuration, $\mathcal{W}(x)$ and $L$ have closed-form expressions. We use the example for in-/anti-plane sliding at the interface of two continuum, for which \cite{Viesca:2016bga}
\begin{equation}
\mathcal{W}(x)=\frac{\sqrt{1-(x/L)^2}}{(1-a/b)\pi/2},\quad \frac{L}{\epsilon L_b} =\frac{1}{\pi(1-a/b)^2}
\end{equation}
and
\begin{equation}
\tilde{\mathcal{L}}(w)=\frac{\epsilon L_b}{2\pi}\int_{-L}^{+L}\frac{\partial w(x,s)/\partial y}{y-x}dy
\label{eq:hilb}
\end{equation}

We look for perturbations of the form
\begin{equation}w(x,s)=\omega(x/L)\exp(\lambda s)\end{equation}
and for shorthand, pass from variable $x/L\Rightarrow x$ such that combining (\ref{eq:redeig})--(\ref{eq:hilb}) leads to the eigenvalue problem
\begin{equation}
\mu\omega(x)=\frac{\sqrt{1-x^2}}{\pi}\int_{-1}^{1}\frac{\omega '(y)}{y-x}dy,\quad \mu=\frac{\frac{a}{b}\lambda-1}{1-a/b}
\end{equation}
The solution to which we find to be, for $n=0,1,2,...$
\begin{equation}
\omega_n(x)=U_n(x)\sqrt{1-x^2},\quad \mu=-(n+1)
\label{eq:eigsoln}
\end{equation}
where $U_n$ is the $n$-th Chebyshev polynomial of the second kind. Solving for $\lambda$, we derive the expression for eigenvalues in the asymptotic limit $a/b\rightarrow 1$
\begin{equation}
\lambda=\frac{1-(n+1)(1-a/b)}{a/b}
\end{equation}
shown as red-dashed curves in Figure 3.

To quickly check the solution (\ref{eq:eigsoln}), we would like to show that
$$-(n+1)U_n(x)=\frac{1}{\pi}\int_{-1}^{1}\frac{\omega'_n(y)}{y-x}dy$$
To do so, we combine the following derivative
$$ \omega'_n(x)=U_n(x)\frac{-x}{\sqrt{1-x^2}}+U'_n(x)\sqrt{1-x^2}$$
 with the identify 
$$U'_n(x)\sqrt{1-x^2}=\frac{-(n+1)T_{n+1}(x)+xU_n(x)}{\sqrt{1-x^2}}$$
to show that
$$\omega'_n(x)=\frac{-(n+1)T_{n+1}(x)}{\sqrt{1-x^2}}$$
where $T_n$ is the $n$-th Chebyshev polynomial of the first kind.
We combine this last result with the property of $T_n(x)$ that \cite{Mason:2003tc}
$$U_{n-1}(x)=\frac{1}{\pi}\int_{-1}^1 \frac{T_n(y)}{\sqrt{1-y^2}}\frac{ds}{y-x}$$
to show that which was to be demonstrated.

\section{Appendix C}
\subsection{C.1 Numerical solutions for quasi-static slip-rate evolution}

We numerically solve for the evolution of the slip rate $V$ and state $\theta$ indirectly by solving instead for the evolution of 
\begin{equation}
v= \ln \frac{V}{V_o} \quad \Theta =\ln\frac{\theta}{\theta_o}
\end{equation}
which remain of comparable order of magnitude during the divergence of slip rate. The evolution equations for $v$ and $\Theta$ with respect to time follow from the time derivative of (\ref{eq:start})
\begin{equation}
\sigma\left(a \frac{\partial v}{\partial t}+b\frac{\partial \Theta}{\partial t}\right)=\frac{\partial\tau_b}{\partial t}+\mathcal{L}[V_o\exp(v)]
\end{equation}
where $\mathcal{L}$ is the second operator in (\ref{eq:L}), reflecting slip between elastic half-spaces, and
\begin{equation}
\frac{\partial \Theta}{\partial t}=\frac{1}{\theta}\frac{\partial\theta}{\partial t}
\end{equation}
with $\partial \theta/\partial t$ given by (3). Owing to the finite-time nature of the instability, integration of the evolution equations with respect to time $t$ is numerically inconvenient as ever-diminishing increments in time are required to resolve order-of-magnitude changes in slip rate. Instead, we make use of that slip also diverges within finite time and integrate with respect to the monotonically increasing slip at the center of the instability $\delta_0=\delta(x=0,t)$, where slip rates are largest.
\begin{equation}
\frac{\partial v}{\partial\delta_0}=\frac{1}{V(0,t)}\frac{\partial v}{\partial t}
\end{equation}
The integration of the evolution equations is done using an adaptive-step scheme with the Hilbert transform operation in $\mathcal{L}$ numerically evaluated using a spectral method \cite{Weideman:1995be, Viesca:2016kua}.

Instability is provoked by a locally applied external force, in the form of a smooth, compact distribution 
\begin{equation}
\frac{\partial\tau_b}{\partial t}=C \left(1-\left(x/L_\tau\right)^2\right)^{3/2}
\end{equation}
For the solutions presented in Figures 1 and 4, $L_\tau=\epsilon L_b$ and $C=\sigma b/(D_c/V_o)$. For values of the parameters $\epsilon$ and $a/b$ for which solutions of the form (2) are stable (i.e., attractive), any memory of initial conditions vanishes as the instability progresses, and the particular choice of initial conditions here---steady state: i.e.,  $v$ and $\Theta=0$ uniformly along $x$---is arbitrary.

\begin{figure}
\includegraphics{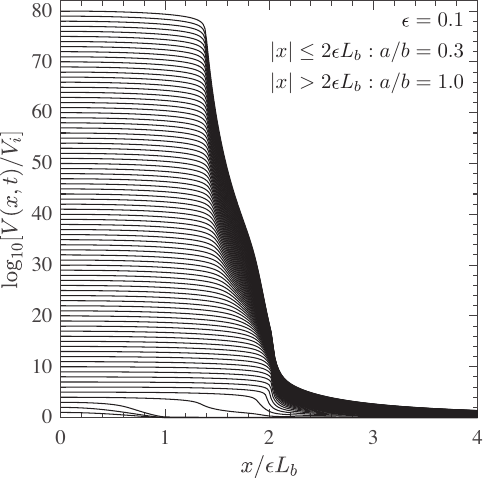}
\caption{\label{suppfig} Snapshots of diverging slip velocity $V$ with distance along the interface $x$. The instability in slip rate occurs within a steady-state rate-weakening patch ($a/b<1$) confined to $|x|\leq 2\epsilon L_b$. For $|x|>2\epsilon L_b$, the frictional response is steady-state rate neutral ($a/b=1$), which alone cannot sustain an instability. The instability is self-sustaining within the rate-weakening patch despite the size of the rate-weakening patch being much smaller than critical wavelength $\lambda_{cr}$ predicted by linear stability analysis.}
\end{figure}

\subsection{C.2 Nucleation on lengthscales below critical length of linear stability analysis}
A slip instability can be provoked under a finite external forcing provided the size of the rate weakening region is greater than the nucleation length $L$. In Figure S1 we show that the numerical solution for slip rate evolution along an interface, in which the rate-weakening region width is small compared to the critical wavelength expected from a linear stability analysis of steady-state sliding. Specifically, $\epsilon=0.1$ and the rate-weakening region ($a/b=0.3$) is restricted to $|x|<w$ for which $w=2\epsilon L_b$. For $|x|>w$ $\epsilon=0.1$ and a steady-state rate-neutral response, $a/b=1$; similar results also occur with rate-strengthening behavior, $a/b>1$, along $|x|>w$; in either case, slip is unconditionally stable in this region and $\lambda_{cr}$ does not exist. For $a/b=0.3$, the critical wavelength from a linear stability analysis of steady-state sliding of an unbounded rate-weakening region is 
$$\lambda_{cr}=2\pi\frac{L_b}{1-a/b}\approx 9 L_b$$
which is more than an order of magnitude larger than the rate-weakening patch size $2w$. Given the disparity between $2w$ and $\lambda_{cr}$, a linear stability analysis would predict that an instability should not be possible as a response to infinitesimal perturbations. However, the non-linear analysis of a spring-block model \cite{Gu:1984ua} and the result that a nucleation $L$ exists for which $L=(1.3374...)\epsilon L_b<w$, both indicate that a self-sustaining instability can be provoked in response to a finite perturbation. This expectaction is born out in the numerical solutions shown, in which a diverging slip rate approaches the expected solution (2).

\end{document}